\documentclass[preprint, secnumarabic, amssymb, superscriptaddress, aps, prl,nobibnotes, longbibliography]{revtex4-2}
\usepackage{amsmath}
\usepackage{amssymb}
\usepackage{amstext}
\usepackage{amsopn}
\usepackage{amsfonts}
\usepackage{amsxtra}
\usepackage[english]{babel}
\usepackage{graphicx}
\usepackage{bm}
\usepackage{multirow}
\usepackage{dcolumn}
\usepackage{color}
\usepackage{textcomp}
\usepackage{upgreek}
\usepackage{textgreek}
\usepackage{setspace}
\usepackage{booktabs}
\usepackage{makecell}
\usepackage{tabularx}
\usepackage{ragged2e}
\usepackage{array} 
\usepackage{tabularray}
\usepackage{threeparttable}
\usepackage{hyperref}
\hypersetup{colorlinks,breaklinks,
            urlcolor=[rgb]{0,0,0.64},
            linkcolor=[rgb]{0,0,0.64},
            citecolor=[rgb]{0,0,0.64}}
\usepackage{nccmath}
\usepackage{amssymb}
\usepackage{graphicx}
\usepackage{bm}
\newcommand{\bS}{{\bf S}}

\newcommand{\rrangle}{{\rangle\!\rangle}}
\newcommand{\llangle}{{\langle\!\langle}}

\setcitestyle{super}

\usepackage{multirow}
\usepackage{makecell}

\usepackage[utf8]{inputenc}

\newcommand{\NJU}{National Laboratory of Solid State Microstructures and Department of Physics, Nanjing University, Nanjing 210093, China}
\newcommand{\MaxPlanck}{Max Planck Institute for the Structure and Dynamics of Matter, Luruper Chaussee 149, 22761 Hamburg, Germany}

\newcommand{\UESTC}{School of Optoelectronic Science and Engineering, University of Electronic Science and Technology of China, Chengdu, 611731 China}

\newcommand{\JPSRC}{Jiangsu Physical Science Research Center, Nanjing University, Nanjing 210093, China}
\newcommand{\FLA}{Initiative for Computational Catalysis, The Flatiron Institute, Simons Foundation, New York City, NY 10010, USA}

\newcommand{\CAS}{Beijing National Laboratory for Condensed Matter Physics, Institute of Physics, Chinese Academy of Sciences, Beijing 100190, China}

\begin{document}
\title{\textbf{Coherent terahertz magnon-phonon three-wave mixing in a layered antiferromagnet}}

\author{Liangyue Li}
\thanks{These authors contributed equally to this work.}
\affiliation{\NJU}

\author{Na Wu}
\thanks{These authors contributed equally to this work.}
\affiliation{\MaxPlanck}

\author{Zhengwang Lin}
\thanks{These authors contributed equally to this work.}
\affiliation{\NJU}

\author{Zefen Li}
\affiliation{\UESTC}

\author{Lixin Liu}
\affiliation{\NJU}

\author{Emil Vinas Boström}
\thanks{Correspondence to: \href{mailto:zhangqi@nju.edu.cn}{zhangqi@nju.edu.cn} and \href{mailto:emil.bostrom@mpsd.mpg.de}{emil.bostrom@mpsd.mpg.de}.}
\affiliation{\MaxPlanck}

\author{Yuan Wan}
\affiliation{\CAS}

\author{Xinbo Wang}
\affiliation{\CAS}

\author{Jianlin Luo}
\affiliation{\CAS}

\author{Fucai Liu}
\affiliation{\UESTC}

\author{Angel Rubio}
\affiliation{\MaxPlanck}
\affiliation{\FLA}

\author{Qi Zhang}
\thanks{Correspondence to: \href{mailto:zhangqi@nju.edu.cn}{zhangqi@nju.edu.cn} and \href{mailto:emil.bostrom@mpsd.mpg.de}{emil.bostrom@mpsd.mpg.de}.}
\affiliation{\NJU}
\affiliation{\JPSRC}

\maketitle

\section{Main Text}

\noindent\textbf{The coherent nonlinear dynamics between collective excitations, such as magnons and phonons, drive emergent phenomena in quantum materials, yet their direct observation remains limited. Here, using double-terahertz-pump optical-probe spectroscopy, we report the direct observation of coherent magnon-phonon three-wave mixing in the layered antiferromagnetic insulator FePS$_{3}$. We resolve both second- and third-order nonlinear responses of antiferromagnetic magnons and identify a suite of nonlinear couplings in two-dimensional (2D) coherent spectra, including definitive sum- and difference-frequency generation between magnons and phonons. These results lay the groundwork for exploiting coherent nonlinearities to entangle magnetic and vibrational excitations, opening avenues for quantum control and hybrid quantum technologies in the terahertz regime.}

%
Collective excitations in solids are fundamental to the dynamical behavior and emergent phenomena of many-body quantum systems\cite{rodin2020collective,xu2025time}. While decades of research have established a detailed understanding of their linear responses\cite{Basov2011RevModPhys}, the nonlinear and nonequilibrium dynamics of these excitations remain far less explored\cite{kimel2007nonthermal,eisert2015quantum,de2021colloquium}. This challenge is particularly acute in low-dimensional and correlated materials, where nonlinear interactions often produce complex phase diagrams and competing quantum orders\cite{stojchevska2014ultrafast,zhang2016cooperative,ilyas2024terahertz,diederich2025exciton}.


Nonlinear processes involving collective excitations encode higher-order interactions and couplings that are inaccessible to linear spectroscopy\cite{cundiff2009optical,wan2019resolving}. Among these, three-wave mixing (TWM) represents a fundamental higher-order interaction, where two waves couple to generate a third at their sum or difference frequency. In magnetic systems, nonlinear interactions rooted in TWM of magnons, such as parametric amplification\cite{kamimaki2020parametric,sheng2023three_mag_coup,sud2025electrically,arfini2025magnon,zhang2025terahertz} and harmonic generations\cite{zhang2024terahertz,huang2024extreme,lan2025coherent}, are not only central to understanding anharmonic interactions in correlated quantum matter, but also critical for developing spin-wave-based information processing\cite{zheng2023tutorial}. This concept extends directly to hybrid excitations, where coherent magnon-phonon three-wave mixing represents the fundamental nonlinear mechanism for spin and lattice subsystems. It provides a powerful tool for manipulating coupled spin and lattice orders beyond equilibrium, yet its direct observation has remained an experimental challenge.


From a quantum perspective, coherent magnon-phonon TWM offers a compelling platform for entangling spin and lattice degrees of freedom\cite{zhang2016cavity,zheng2023tutorial}. Analogous to quantum optics\cite{braunstein2005quantum}, difference-frequency generation (DFG) is an essential mechanism for generating nonclassical states\cite{wu1986generation,kwiat1995new}, such as squeezed magnons and phonons, which are foundational resources for continuous-variable quantum information\cite{li2018magnon,li2019squeezed,chakraborty2023nonreciprocal}. Hybrid magnon-phonon entanglement could thus enable the integration of magnetic and vibrational excitations into hybrid quantum architectures, even at terahertz (THz) frequencies.


Despite its significance, the direct observation of coherent magnon-phonon TWM has been elusive. Recent advances in time-domain nonlinear spectroscopy with ultrafast THz pulses have enabled the generation and detection of nonlinear signals in solids, including magnon harmonic generation in rare-earth orthoferrites\cite{lu2017coherent,zhang2023generation,huang2024extreme,zhang2024terahertz,zhang2025spin}, phonon nonlinearities in complex oxides\cite{forst2011nonlinear} and topological insulators\cite{blank2023two}, and signatures of photon-mediated magnon-phonon coupling in CoF$_{2}$\cite{mashkovich2021terahertz,metzger2024magnon}. However, the elemental TWM processes between magnons and phonons, namely sum-frequency generation (SFG) and DFG, have not yet been coherently and directly resolved.
\par

In this work, we present the direct observation of coherent nonlinear dynamics of magnons and phonons in the layered antiferromagnetic (AFM) insulator FePS$_{3}$, with double-THz-pump optical-probe 2D coherent spectroscopy. Both second- and third-order nonlinear magnonic processes are identified. The 2D coherent spectra further reveal a rich landscape of nonlinear magnon-phonon interactions, including TWM processes such as sum- and difference-frequency generation. Theoretically, we develop a Liouville-space Lindblad framework to model the coherent nonlinear magnon-phonon dynamics, which well reproduces the THz 2D spectra. Our findings establish a foundation for harnessing coherent nonlinearities to generate entanglement and to develop advanced quantum control schemes across spin and lattice degrees of freedom in antiferromagnetic systems.

\begin{figure*}[tb]
\includegraphics[width=\textwidth]{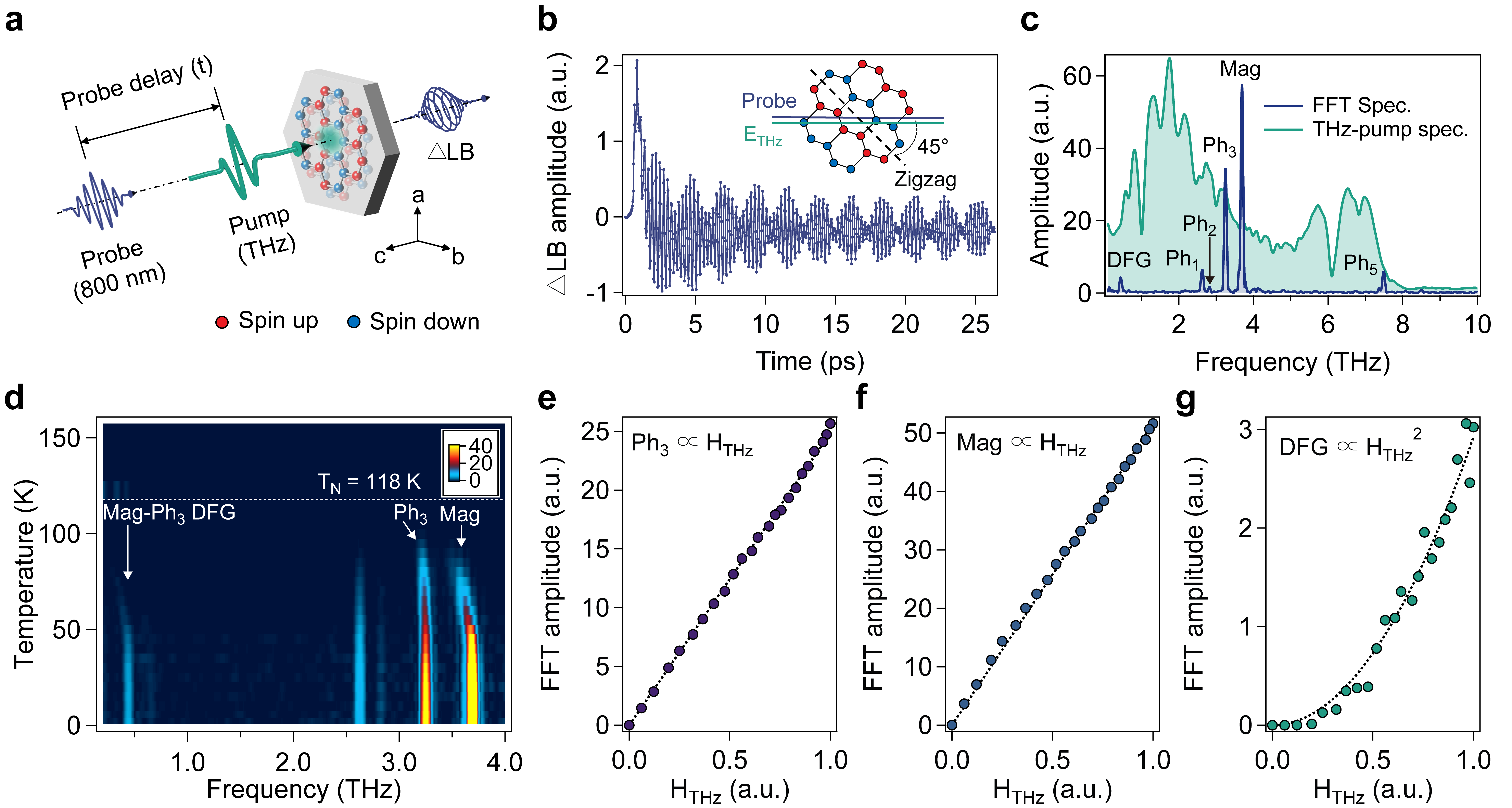} 
\caption{\textbf{Intense THz-pump optical-probe coherent spectra of FePS$_{3}$ at 0~T.} \textbf{a}, Schematic diagram of experimental geometry. THz-induced coherent oscillations are observed through the ellipticity change of 800 nm probe pulses. \textbf{b}, Coherent phonons and magnons in a 20~\textmu m-thick FePS$_{3}$ single crystal, driven by THz pulses of 2.5 MV/cm in strength. The insert illustrates the polarization direction of the THz and the probe light relative to the zigzag spin chain of the FePS$_{3}$. \textbf{c}, Spectra of the THz-pump pulse and the THz-induced coherent oscillations. Doubly degenerated AFM magnons (Mag) are located at 3.7~THz. \textbf{d}, Frequency-domain pump-probe spectra from 2.2 K to 150 K. \textbf{e–g}, Amplitude of Ph$_{3}$, Mag, and Mag-Ph$_{3}$ difference-frequency generation (DFG) as a function of the THz-field strength. The magnon and Ph$_{3}$ scale linearly with the THz field, while the Mag-Ph$_{3}$ DFG scales quadratically. We note that Mag and Ph$_{3}$ are driven by the magnetic field component ($H_\text{THz}$) of the THz pulse, and the corresponding THz electric-field strength ($E_\text{THz}$) is directly measured and normalized by a field strength of 2.5 MV/cm. }  
\label{fig1}
\end{figure*}
\par
FePS$_{3}$, a van der Waals AFM insulator, emerges as an ideal material platform to explore these nonlinear dynamics due to its prominent spin-lattice interactions\cite{zhou2022dynamical,zong2023spin}. It belongs to the family of transition metal phosphorus trichalcogenides, with magnetic Fe ions forming a honeycomb lattice within each layer\cite{lanccon2016magnetic,lee2016ising}. Below its Néel temperature (T\textsubscript{N} = 118 K), FePS$_{3}$ develops a zigzag AFM order with spins oriented out-of-plane, and has space group symmetry C2/m\cite{lee2016ising}. The zigzag order breaks in-plane rotational symmetry, enabling linear dichroism (LD) and birefringence (LB) associated with the magnetic order\cite{zhang2021spin,zhang2021observation}—properties crucial for optical probing of coherent collective excitations. Furthermore, the enlarged primitive magnetic unit cell folds zone-boundary (M-point) optical phonons to the Γ point, yielding modes at 2.64 THz (Ph$_{1}$), 2.83 THz (Ph$_{2}$), and 3.25 THz (Ph$_{3}$)\cite{lee2016ising}. The linear hybridization of these phonons with AFM magnons, forming magnon polarons, has been observed via static Raman and THz spectroscopy\cite{liu2021direct,zhang2021coherent,vaclavkova2021magnon,cui2023chirality}. Recent demonstrations of coherent magnon polaron manipulation using intense THz pulses\cite{luo2025terahertz}, and the induction of a metastable ferromagnetic phase\cite{ilyas2024terahertz}, underscore the strong, complex spin-lattice interactions in this system\cite{mertens2023ultrafast,ergeccen2023coherent}. These established properties make FePS$_{3}$ a particularly promising candidate for exploring the coherent nonlinear magnon-phonon processes central to this work.
\par
\begin{figure*}[tb]
\includegraphics[width=\textwidth]{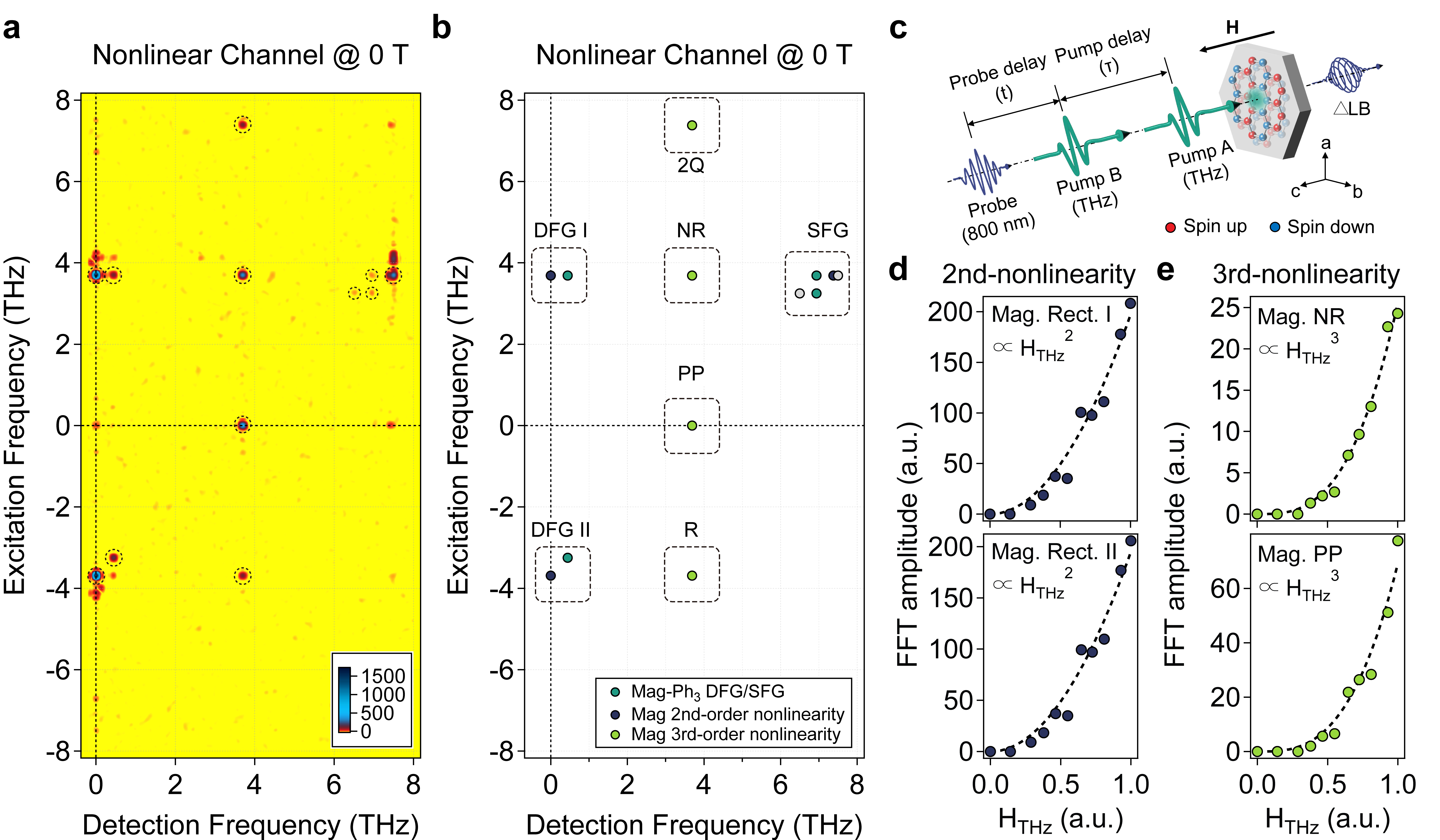} 
\caption{\textbf{Double-THz-pump optical-probe 2D coherent spectra of FePS$_{3}$ at 0~T.} \textbf{a}, Nonlinear channel of the 2D spectra. Strong magnon-magnon and magnon-phonon nonlinear features are observed and circled with dashed lines. \textbf{b}, The labeling of nonlinear signals shown in \textbf{a}. Non-rephasing (NR), rephasing (R), pump-probe (PP), double-quantum (2Q), DFG and sum-frequency generation (SFG) regions in the 2D spectra are marked. \textbf{c}, Schematic diagram of experimental geometry of 2D coherent spectra with external magnetic field. \textbf{d}, THz-Field dependence of magnon 2nd-order nonlinear processes (Rectification) in the DFG I and DFG II region. \textbf{e}, THz-field dependence of magnon 3rd-order nonlinear processes (NR and PP). The THz electric-field strength is measured and normalized to 1.1 MV/cm.}
\label{fig2}
\end{figure*}

To investigate the nonlinear dynamics of collective excitations and identify their quantum pathways, we performed single- and double-THz-pump optical-probe coherent spectroscopy on a 20~\textmu m-thick FePS$_{3}$ single crystal. As illustrated in Fig.~\ref{fig1}a, an intense THz pulse with a field strength up to 2.5~MV/cm impinges normally on the FePS$_{3}$ flake, see Supplementary Note 2. A linearly polarized 800~nm probe pulse, delayed by time $t$, is transmitted through the sample. Owing to the strong LB associated with the AFM order, THz-induced free induction decay of coherent magnons modulates the AFM order parameter, encoding its amplitude and phase information onto the polarization state of the probe light ($\Delta$LB)\cite{ilyas2024terahertz}. Consequently, the $\Delta$LB signal is directly proportional to the amplitude of the AFM magnon, as well as the amplitudes of phonons that hybridize with it (see Supplementary Note 1). Figure~\ref{fig1}b presents a representative one-dimensional (1D) time-domain coherent oscillation of magnon and phonons at 2.2 K, the corresponding spectrum (blue line) is depicted in Fig.~\ref{fig1}c. The 3.7~THz AFM magnons (Mag) dominate the spectral amplitude under the pump-probe polarization scheme shown in the insert of Fig.~\ref{fig1}b, while the nearby modes are all Raman-active phonons (Ph$_{1}$, Ph$_{2}$, Ph$_{3}$, Ph$_{5}$) as previously reported\cite{liu2021direct}. 
\par

In addition to the fundamental modes of the magnon and phonons, remarkably, we observed a distinct mode at 0.45~THz, which precisely matches the frequency difference between the Magnon ($\omega_{\rm Mag}$ = 3.7~THz) and the Ph$_{3}$ phonon ($\omega_{\rm Ph_3}$ = 3.25~THz), as shown in Fig.~\ref{fig1}c. 
This mode is absent from both static THz and Raman spectra, pointing to the emergence of a coherent Mag-Ph$_{3}$ difference-frequency state. 
This interpretation is corroborated by the temperature dependence of the spectra displayed in Fig.~\ref{fig1}d. 
As the temperature increases, the DFG mode gradually redshifts, tracking the softening of the magnon mode while consistently locking to the frequency $\omega_{\rm Mag} - \omega_{\rm Ph_3}$.    

To confirm that the observed magnon-phonon DFG signal originates from the intrinsic nonlinear coupling between collective excitations in FePS$_{3}$, rather than an artifact from the nonlinearity of the optical detection, we investigate the THz-field dependence of both the fundamental and the DFG modes at 2.2 K.
The AFM magnon is driven by the magnetic field component $H_\text{THz}$ of the incident THz pulse. 
Owing to the linear hybridization between the magnon and the Raman-active Ph$_{3}$ mode, the amplitudes of both modes scale linearly with the THz driving field, as shown in Fig.~\ref{fig1}e and f. 
It is worth noting that the magnon and Ph$_{3}$ mode are directly driven by $H_\text{THz}$, the corresponding electric field component $E_\text{THz}$ is measured and normalized to a field strength of 2.5 MV/cm. 
The linear response of these fundamental modes ensures that subsequent nonlinear features are intrinsic to the sample.
In contrast, the amplitude of Mag-Ph$_{3}$ DFG signal exhibits a quadratic dependence on the THz-field strength (Fig.~\ref{fig1}g), establishing it as a second-order nonlinear process involving the collective excitations in FePS$_{3}$.
Together, the low-energy THz photons, the absence of any broadband carrier response, the long-lived coherent oscillations, and the linear field dependence of the infrared-active modes rule out strong electric-field-driven nonlinear effects such as multiphoton absorption, impact ionization, or Zener tunnelling.

To resolve the quantum pathways underlying the Mag-Ph$_{3}$ DFG signal and the accompanying nonlinear processes, we perform two-dimensional (2D) coherent spectroscopy in a double-THz pump-probe configuration. 
As illustrated in Fig.~\ref{fig2}c, two nearly identical intense THz pump pulses A and B are separated by a variable inter-pump delay time $\tau$. 
By systematically scanning $\tau$ together with the probe delay $t$, we acquired four distinct 2D time-domain datasets $S(\tau,t)$, from which the nonlinear contribution $S\textsubscript{NL}$ generated by the joint action of both pumps is isolated as 
$S\textsubscript{NL} = S\textsubscript{AB} - S\textsubscript{A} - S\textsubscript{B}$, where $S\textsubscript{AB}$, $S\textsubscript{A}$, and $S\textsubscript{B}$ denote the 2D time-domain signal recorded with both pump pulses, with pump pulse A only, and with pump pulse B only, respectively (Extended data Fig.~\ref{S:3}).
Figure~\ref{fig2}a presents the resulting nonlinear channel of the 2D spectra at zero magnetic field, obtained by a two-dimensional Fourier transform of $S\textsubscript{NL}$ along $\tau$ and $t$.
The vertical and horizontal axes represent the excitation and detection frequencies, respectively, and spectral features are labeled by the coordinates ($\omega_{\rm Det}$, $\omega_{\rm Exc}$)
in units of THz. 

The nonlinear channel exhibits pronounced signatures associated with the AFM magnon, the Ph$_{3}$ and Ph$_{5}$ phonons, as highlighted by the dashed circles in Fig.~\ref{fig2}a.
Under THz driving, the AFM magnon exhibits second-order nonlinear response, including magnon rectification at $(0, \omega_{\rm Mag})$ and second-harmonic generation (SHG) at $(2\omega_{\rm Mag}, \omega_{\rm Mag})$, while the third-order nonlinear responses comprise double-quantum coherence (2Q) at $(\omega_{\rm Mag}, 2\omega_{\rm Mag})$, non-rephasing (NR) at $(\omega_{\rm Mag}, \omega_{\rm Mag})$, pump-probe (PP) at $(\omega_{\rm Mag}, 0)$, and rephasing (R) signals at $(\omega_{\rm Mag}, -\omega_{\rm Mag})$.
These nonlinear features are labeled in Fig.~\ref{fig2}b. 
THz field-dependent 2D spectral measurements further confirm their nonlinear order: the second- and third-order magnonic signals scale quadratically (Fig.~\ref{fig2}d) and cubically (Fig.~\ref{fig2}e) with THz-field strength, respectively. 
The origin of these second magnetic signals in FePS$_3$ differs fundamentally from that reported in rare-earth orthoferrites\cite{lu2017coherent,zhang2023generation,zhang2024terahertz}, where the Dzyaloshinskii–Moriya interaction (DMI) plays a central role. In FePS$_{3}$, by contrast, DMI is absent, and the spin precession nominally preserves rotational symmetry along the z-axis that would otherwise suppress second-order magnonic responses. Their observation, therefore, implies the presence of an additional symmetry-breaking mechanism, which we attribute to nonlinear magnon-phonon coupling.

In addition to the purely magnonic nonlinearities, 2D spectrum reveals clear signatures of magnon-phonon TWM. 
The sum-frequency generation (SFG) feature between the magnon and Ph$_{3}$ mode appears at $\big(\omega_{\rm Mag}+\omega_{\rm Ph_3}, \omega_{\rm Ph_3}\big)$ and $\big(\omega_{\rm Mag}+\omega_{\rm Ph_3}, \omega_{\rm Mag}\big)$, while two Mag-Ph$_{3}$ DFG features are resolved at $\big(\omega_{\rm Mag}-\omega_{\rm Ph_3}, \omega_{\rm Mag}\big)$ and $\big(\omega_{\rm Mag}-\omega_{\rm Ph_3}, -\omega_{\rm Ph_3}\big)$. 
Together, these DFG channels account for the difference-frequency mode observed in the single-pump spectra.
The remaining nonlinear peaks in Fig.~\ref{fig2}a originate from photon-mediated magnon-phonon up-conversion and phonon-phonon processes. A complete assignment of all nonlinear features in the 2D spectrum is provided in Supplementary Table S2.

\par

\begin{figure*}[tb]
\includegraphics[width=\textwidth]{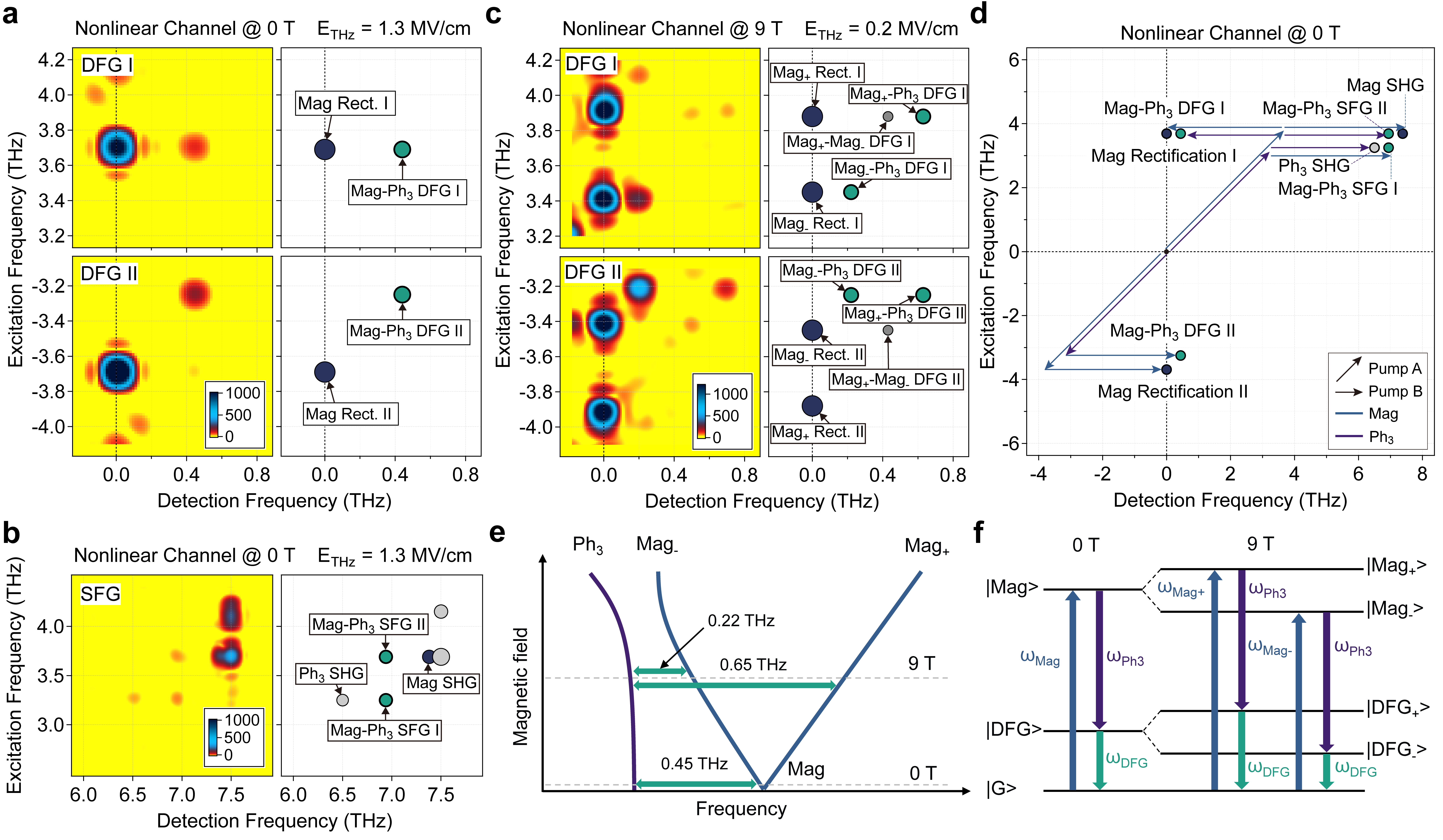} 
\caption{\textbf{Coherent nonlinear magnon-phonon coupling in FePS$_{3}$ revealed by the 2D spectra.} \textbf{a, b}, The DFG (\textbf{a}) and the SFG (\textbf{b}) regions of the nonlinear channel 2D spectra at 0~T. The left panel in each subfigure is an experimental 2D spectrum, while the nonlinear processes are labelled in the right panel. \textbf{c}, Nonlinear channel 2D spectra in the DFG I and DFG II regions at 9~T. Zeeman splitting of Mag Rectification and Mag-Ph$_{3}$ DFG features are observed. \textbf{d}, Frequency-vector diagram of Mag Rectification/SHG and Mag-Ph$_{3}$ DFG/SFG processes at 0~T. \textbf{e}, Schematic diagram of Ph$_{3}$ and Mag frequency under magnetic fields \cite{zhang2021coherent}. The green arrows marked the energy difference between AFM magnons and Ph$_{3}$ phonon. \textbf{f}, Schematic energy-level diagram of Mag-Ph$_{3}$ DFG processes at 0~T and 9~T.}
\label{fig3}
\end{figure*}
\par   
We now turn to a closer inspection of the DFG regions in the nonlinear channel 2D spectrum.
Figure~\ref{fig3}a presents a zoom-in version of DFG I and DFG II regions at 0 T. 
The DFG I region is located at the non-rephasing quadrant of the 2D spectrum, defined by a positive excitation frequency $\omega_{\rm Exc}$. 
Within this quadrant, a prominent Mag-Ph$_{3}$ DFG feature appears at (0.45, 3.7), corresponding to $\big( \omega_{\rm Mag}-\omega_{\rm Ph_3},\, \omega_{\rm Mag} \big)$, immediately adjacent to the magnon-rectification peak at (0, 3.7). 
The associated quantum pathway proceeds as follows: pulse A first prepares a coherent AFM magnon state, which is subsequently coupled by pulse B to a Ph$_{3}$ phonon carrying the opposite phase relative to the magnon, and the resulting hybrid mode evolves at the difference frequency $\omega_{\rm Mag}-\omega_{\rm Ph_3}$.
The conjugated DFG process is identified at ($0.45, -3.25$) in the DFG II region, located in the rephasing quadrant where the negative $\omega_{\rm Exc}$ encodes a negative-phase coherence.
This peak corresponds to $\big(\omega_{\rm Mag}-\omega_{\rm Ph_3}, -\omega_{\rm Ph_3}\big)$ and is generated by the reversed pathway: pump pulse A first excites a single-phonon coherence with a negative phase, after which pulse B drives the magnon with a positive phase, again yielding a hybrid coherence at the difference frequency.
The corresponding quantum pathways of both DFG processes are illustrated as frequency-vector diagrams shown in Fig.~\ref{fig3}d. 

\par

To further investigate magnon-phonon nonlinear dynamics, we apply an external magnetic field of up to 9 T along the $c$-axis of FePS$_{3}$. 
The corresponding nonlinear signals in DFG I and II regions are shown in Fig.~\ref{fig3}c. 
At 9~T, the AFM magnon splits into two branches Mag$_{+}$ and Mag$_{-}$ (Fig.~\ref{fig3}e), and the magnon-rectification peak splits accordingly into rectification signals of Mag$_{+}$ and Mag$_{-}$, respectively. 
Meanwhile, the Mag-Ph$_{3}$ DFG I features at (0.45, 3.7) also bifurcates into Mag$_{-}$-Ph$_{3}$ DFG I at (0.22, 3.45) and a Mag$_{+}$-Ph$_{3}$ DFG I peak at (0.65, 3.88). 
An analogous splitting is observed for DFG II: the original peak at (0.45, -3.25) splits into Mag$_{-}$-Ph$_{3}$ and Mag$_{+}$-Ph$_{3}$ features at (0.22, -3.23) and (0.65, -3.23), respectively. The positions of all Mag-Ph$_{3}$ DFG peaks agree quantitatively with the static magneto-THz measurements and energy-level diagram in Fig.~\ref{fig3}e and f.
In addition, a nonlinear Mag$_{+}$-Mag$_{-}$ DFG signal also emerges (grey circles in Fig.~\ref{fig3}c), reminiscent of the magnon-pair difference-frequency response recently reported in YFeO$_{3}$\cite{zhang2024terahertz}.

In comparison with the DFG processes, the Mag-Ph$_{3}$ SFG signals are markedly weaker in the 2D spectrum. 
Figure~\ref{fig3}b shows the SFG features resolved under intense THz driving at 1.3 MV/cm. 
Two peaks are clearly identified at (6.95, 3.25) and (6.95, 3.7), which we assign to Mag-Ph$_{3}$ SFG I and II, respectively.
The detection frequency of 6.95 THz coincides with the sum frequency of Mag and Ph$_{3}$. 
The two SFG processes can be described as follows: pump pulse A first prepares a coherent Mag (Ph$_{3}$) state, after which pulse B drives the complementary Ph$_{3}$ (Mag) excitation, generating a hybrid coherence whose phase accumulates at the sum frequency.
These features are not resolved in the 1D measurements and do not appear clearly in the magnetic field measurements due to the reduced THz field inside the magnet (see Supplementary Note 2). The peak at (6.5, 3.25) corresponds to second-harmonic generation of Ph$_{3}$, while the feature at (7.5, 4.1) originates from a phonon-phonon nonlinearity that lies beyond the scope of the present work.
Notably, no pronounced TWM signals are observed between the magnon and the other magnon-polaron modes (Ph$_{1}$ and Ph$_{2}$).
\par

\begin{figure*}[tb]
    \centering
    \includegraphics[width=\textwidth]{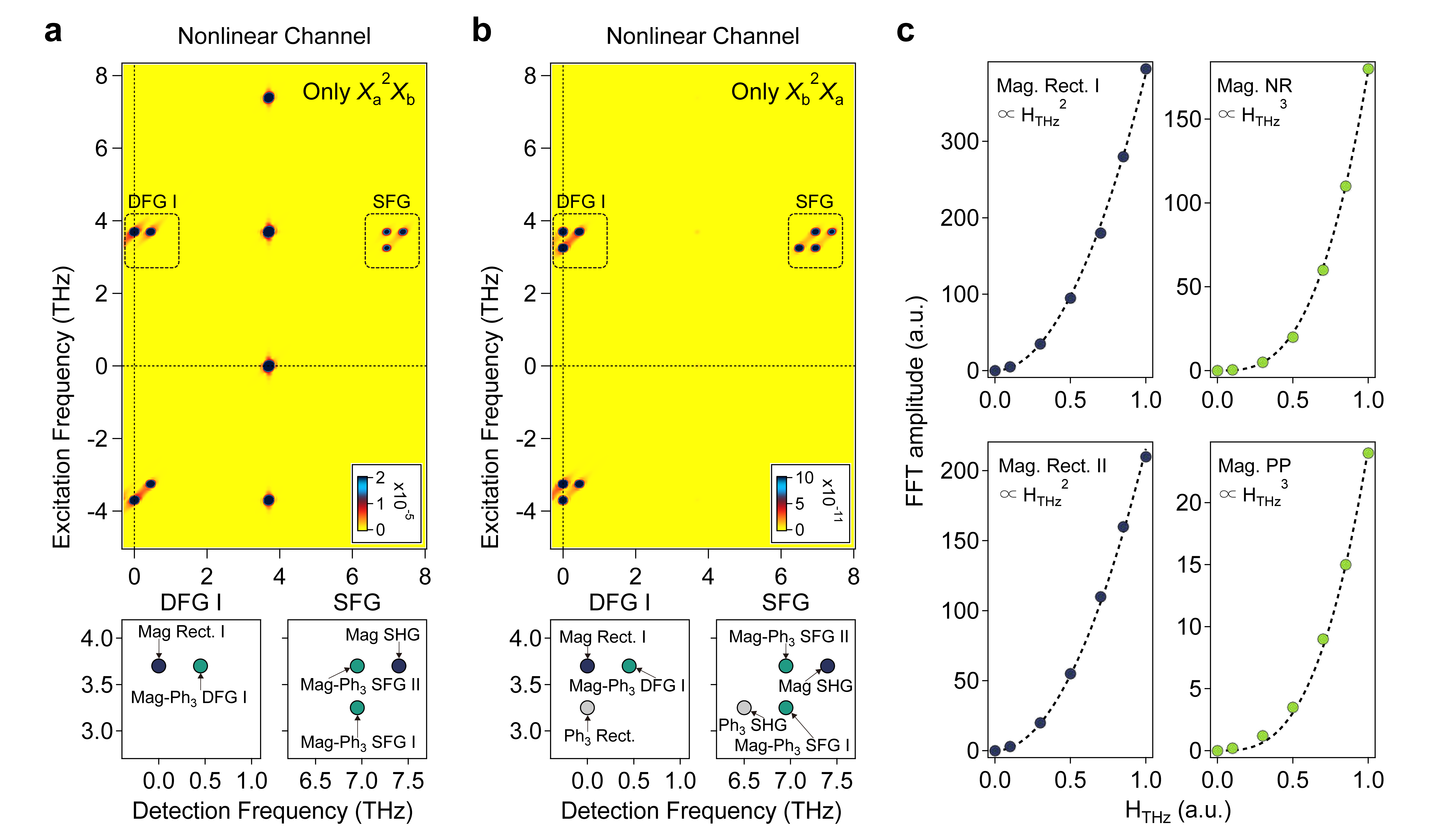}
    \caption{\textbf{Theoretical simulation of nonlinear magnon-phonon features in the 2D spectra.}
    \textbf{a, b}, The calculated nonlinear channel of the THz 2D coherent spectra, including only the $X_a^2 X_b$ term (\textbf{a}), and only the $X_b^2 X_a$ term (\textbf{b}) as nonlinearities. The coupling parameter $\beta_{aa}^b = 17$~meV/{\AA}, as determined by first-principle calculations~\cite{ilyas2024terahertz}, and $\beta_{bb}^a = 0.017$~meV/{\AA}$^2$. 
    The $X_a^2 X_b$ coupling captures the essential magnon-phonon TWM signatures, while the $X_b^2 X_a$ channel provides distinct Ph$_{3}$ Rectification peaks that are not observed in the experiment.
    \textbf{c}, the THz-field dependence of the simulated nonlinear features.}
    \label{fig4}
\end{figure*}

The observed nonlinear magnonic and magnon-phonon TWM signals cannot be captured by linear spin-wave theory or by the linear magnon-phonon hybridization, both of which are restricted to quadratic forms of bosonic operators. 
Nonlinear magnon-magnon interactions arise naturally from higher-order terms in the Holstein–Primakoff expansion. Meanwhile, the Mag-Ph$_3$ DFG and SFG processes, corresponding to TWM signals, require cubic coupling terms of magnon-phonon bosonic operators in the Hamiltonian. 
Although these high-order interactions remain hidden in the equilibrium spectrum and under weak excitation, they become directly accessible in the strong-field regime probed by 2D THz spectroscopy.

To capture the nonlinear dynamics among the collective excitations, we start from the microscopic Hamiltonian of FePS$_{3}$ presented in the Methods.
At zero magnetic field, the magnon branch is doubly degenerate; since the analysis below is insensitive to the internal structure of the degeneracy, we treat the magnon doublet as an effective single mode with frequency $\omega_{\rm Mag}$. 
Restricting our attention to the DFG and SFG signal at 0~T and retaining only the Ph$_3$ phonon, the effective Hamiltonian expanded up to cubic order in the bosonic operators, takes the form as
\begin{align}\label{eq:total}
    \mathcal{H} &= \omega_{\rm Mag} a^\dagger a + \omega_{\rm Ph_3} b^\dagger b + gX_aX_b\\
    &+ \beta_{aa}^b X_a^2 X_b + \beta_{bb}^a X_b^2 X_a 
     + \mu_a X_a H(t). \nonumber
\end{align}
where $a$ and $b$ denote the annihilation operators of the magnon and Ph$_3$ phonon, with frequencies $\omega_{\rm Mag} = 3.7$~THz and $\omega_{\rm Ph_3} = 3.25$~THz, and $X_a = a^{\dagger}+a, X_b=b^{\dagger}+b$ denote the corresponding displacement quadratures.
The first two terms in the Hamiltonian describe the free Hamiltonian of the two modes, the third captures their linear hybridization with coupling strength $g$, and the fourth and fifth terms introduce the cubic magnon-phonon couplings $\beta_{aa}^b$ and $\beta_{bb}^a$ that drive the TWM processes.
The system is excited by the magnetic-field component $H(t)$ of the THz pulses, which couples to the magnon through the effective magnetic moment $\mu_a$. 
Although no explicit magnon Kerr term is included in the Hamiltonian, our simulations naturally capture cascaded nonlinear processes generated by multi-pulse excitations: successive actions of the cubic TWM terms can produce effective third-order responses without invoking an intrinsic Kerr nonlinearity.

Microscopically, the two nonlinear couplings $\beta_{aa}^b X_a^2 X_b$ and $\beta_{bb}^a X_b^2 X_a$ originate from distinct terms of the spin-phonon Hamiltonian.
The former, quadratic in the magnon operators, is generated by the phonon modulation of the dominant magnetic exchange interaction in FePS$_3$, and has been calculated from first-principles calculations to be $\beta_{aa}^b \approx 17$~meV/{\AA}~\cite{ilyas2024terahertz}. The latter, in contrast, can only arise from single-ion or symmetric exchange anisotropies, since the DMI are forbidden by the crystal symmetry of FePS$_3$. 
Because such anisotropies scale with the spin-orbit coupling, which is weak in this system, $\beta_{bb}^a$ is suppressed relative to $\beta_{aa}^b$ by a factor $\alpha_{\rm SOC} \sim 0.1$.
Furthermore, being quadratic in phonon displacement, the $\beta_{bb}^a$ term acquires an additional smaller factor $Q_b/l$, where $Q_b$ is the phonon displacement amplitude and $l$ is the oscillator length.
Combining these two contributions yields the estimate $\beta_{bb}^a \sim 0.001\beta_{aa}^b$, establishing the two-magnon-one-phonon channel as the dominant cubic nonlinearity in our system.

To model the coherent nonlinear dynamics of coupled magnons and phonons and simulate the 2D THz spectra, we develop a Liouville-space Lindblad framework~\cite{reitz_nonlinear_2025, Mukamel1986,dorfman_2016}. 
This approach provides a unified theoretical description in which dissipation is incorporated from the outset and the nonlinear pathways are explicitly resolved, enabling a systematic extraction of the second- and third-order response functions probed in 2D THz spectroscopy (see Methods for details).
The framework reproduces all qualitative features of the experimental 2D spectrum at the phenomenological level.
Figure~\ref{fig4} presents the simulated coherent 2D THz spectrum obtained from the two-mode model of Eq.~\ref{eq:total}, at zero magnetic field (see Methods). 
To isolate the dominant nonlinear coupling channel in FePS$_{3}$, we first compute the spectrum with the $\beta_{aa}^b X_a^2 X_b$ term as the sole source of nonlinear mode mixing, i.e., setting $\beta_{bb}^a = 0$.
As shown in Fig.~\ref{fig4}a, the prominent Mag-Ph$_3$ TWM features are reproduced: DFG peaks emerge at $\big( \omega_{\rm Mag}-\omega_{\rm Ph_3},\, \omega_{\rm Mag} \big)$ and $\big( \omega_{\rm Mag}-\omega_{\rm Ph_3},\, -\omega_{\rm Ph_3} \big)$, while SFG peaks occur at $\big(\omega_{\rm Mag}+\omega_{\rm Ph_3},\, \omega_{\rm Ph_3}\big)$ and $\big(\omega_{\rm Mag}+\omega_{\rm Ph_3},\, \omega_{\rm Mag}\big)$. 
The simulation further reproduces the second-order (Rectification and SHG) and third-order (PP, 2Q, NR, and R) magnonic responses.
The agreement between simulation and experiment confirms that the model captures the essential physics of magnon-phonon TWM. 
The calculated field dependence of the nonlinear magnon processes is also consistent with the experimental observations, as shown in Fig.~\ref{fig4}c.

In contrast, the spectrum generated by the $\beta_{bb}^a X_b^2 X_a$ coupling alone (with $\beta_{aa}^b = 0$) differs markedly, as shown in Fig.~\ref{fig4}b.
While this coupling still produces the Mag-Ph$_{3}$ DFG and SFG features, their amplitudes are far smaller than those obtained from the $X_a^2 X_b$ channel.
Additionally, the $X_b^2 X_a$ coupling yields Ph$_{3}$ rectification peaks that are absent from the experimental spectrum.
It also yields a Ph$_{3}$ SHG peak, but this feature may also originate from phonon-phonon nonlinearities not included in the simulation.
The comparison between simulation and experiment thus establishes the two-magnon-one-phonon coupling $X_a^2 X_b$ as the dominant TWM mechanism in FePS$_{3}$.


Within our theoretical framework, only the Mag-Ph$_{3}$ coupling term is derived from first principles and quantitatively parameterized, yet the minimal input is already sufficient to qualitatively reproduce the full set of TWM and nonlinear-magnon features observed in experiment. 
A fully quantitative description, although feasible in principle, would require both systematic accounting of multiple experimental conditions and an explicit treatment of all relevant collective modes together with their nonlinear couplings, a comprehensive undertaking that lies beyond the scope of the present work.

The coherent magnon-phonon TWM also uncovers a direct route to generating hybrid non-classical states of these excitations. 
In a DFG process, a strong pump excitation can decay into, or coherently couple two lower-energy modes, thereby effectively realizing a parametric amplification mechanism.
Under strong driving, the linearized TWM Hamiltonian acquires the canonical form $(a^\dagger b^\dagger+ab)$, which corresponds to a two-mode squeezing operation, a fundamental quantum primitive that can generate entangled and squeezed magnon-phonon states.
Such states constitute essential resources for continuous-variable quantum information processing.
As a proof-of-principle, we compute the squeezing of the magnon and Ph$_{3}$ mode induced by a broadband THz pulse in the presence of the magnon-phonon nonlinearities (Extended data Fig.~\ref{S:7}).
Although the effect remains modest for current experimental parameters, the calculations confirm that optimal two-mode squeezing is achievable and could be substantially enhanced with narrowband THz sources.

These observations establish a layered antiferromagnetic system with intrinsic nonlinear couplings as a natural platform for cavity magnomechanics in the THz regime.
The well-established Yttrium Iron Garnet (YIG) cavity magnomechanical systems operate in the microwave range, combining GHz magnons with much lower-frequency acoustic phonons of tens of MHz within a GHz cavity\cite{Tabuchi2014PRL,zhang2016cavity}. 
The FePS$_3$ system instead hosts THz magnons and optical phonons of comparable energies, together with a giant intrinsic linear magnon-phonon coupling strength of $\sim83$ GHz, a cooperativity of 15\cite{zhang2021coherent}, and the strong two-magnon-one-phonon nonlinearity uncovered here.
The van der Waals nature of 2D antiferromagnets further facilitates seamless integration into THz cavities and on-chip architectures\cite{kipp2025cavity}.
Layered antiferromagnet-based THz cavity magnomechanical platforms could therefore open new avenues for ultrafast quantum control and transduction on picosecond time scales.
More broadly, the rich collective excitations of layered quantum materials and their heterostructures provide a versatile testbed for quantum-state engineering, paving the way toward high-speed, miniaturized quantum devices that exploit the strong intrinsic nonlinearities of these emergent platforms.
\par

\bibliographystyle{apsrev}
\bibliography{Ref}
\newpage
\section{Methods\label{sec:methods}}
\noindent\textbf{Single crystal synthesis.} Single crystals of FePS$_{3}$ were synthesized by the chemical vapor transport (CVT) method using iodine as the transport agent. Stoichiometric amounts of Iron powder (99.998 \%), phosphorus powder (98.9 \%), and sulfur powder (99.9995 \%) were mixed with iodine (1 mg/cc) and sealed under high vacuum in 25 cm-long quartz tubes. The tubes were placed in a horizontal one-zone tube furnace with the charge positioned near the center. Sizeable crystals (10×10×0.5 mm$^3$) were obtained by gradually heating the precursor to 750 °C, holding for a week, and cooling to room temperature. A 20~\textmu m-thick flake was mechanically exfoliated from the single crystal along [0,0,1] direction.\\

\noindent\textbf{Single and double-THz pump optical-probe coherent spectroscopy.} The experimental setup utilizes a 35~fs, 1~KHz regen-amplifier (Spectra-Physics Spitfire Ace) as the primary light source, which drives an optical parametric amplifier (OPA, Light Conversion TOPAS-Prime). The OPA generates 1300 nm pump light (0.48 mJ), which is split into two beams of equal intensity (0.22 mJ per arm). These beams are directed onto the organic crystal DSTMS (4-N, N-dimethylamino-4’-N’-methyl-stilbazolium 2, 4, 6-trimethylbenzenesulfonate) to produce double-THz pump pulses with peak electric fields up to 1.3 MV/cm. In the single-THz pump configuration, the THz electric field can reach up to 2.5 MV/cm. Samples are mounted in a (custom-built) low-vibration closed-cycle cryostat (2.2–300 K), with both the pump and probe pulses normally incident on the sample. The THz-induced polarization change in the transmitted probe light is detected using a balanced detection scheme. The entire THz beam path is purged with dry air to minimize absorption and scattering. Two-dimensional spectroscopy under magnetic fields was performed at the Synergetic Extreme Condition User Facility (SECUF). A liquid-helium-cooled superconducting magnet (Oxford Spectromag-10T) was employed, with the sample placed inside a variable-temperature insert. The magnetic field was applied in the Faraday geometry, aligned with both the sample's spin direction and the direction of light propagation. In this setup, the $E_\text{THz}$ at the sample position inside the magnet can reach up to 200 KV/cm.
\\

\noindent\textbf{Microscopic spin-phonon Hamiltonian.}
The spin-phonon coupling in FePS$_3$ is described by a Hamiltonian of the form~\cite{cui2023chirality}
\begin{align}\label{eq:ham}
 \mathcal{H} &= \sum_{ij} \bS_i \cdot ({\bf J}_{ij}\,\bS_j) - \Delta \sum_i (S_i^z)^2 \nonumber\\
 &\quad+ \frac{1}{2} \sum_\alpha \bigg[ P_\alpha^2 + \Omega_\alpha^2\, Q_\alpha^2 \bigg],
\end{align}
where the tensor ${\bf J}_{ij}$ encodes the (possibly anisotropic) exchange coupling between the spins $\bS_i$ and $\bS_j$, and $\Delta$ parameterizes the single-ion anisotropy. 
The operators $P_\alpha$ and $Q_\alpha$ denote the momentum and displacement of phonon mode $\alpha$, respectively, and $\Omega_\alpha$ is the corresponding bare phonon frequency. 
Since the magnetic interactions depend on the instantaneous atomic positions, they give rise to a coupling interaction between the spins and the lattice displacements.
For small displacements $Q_\alpha$, the exchange tensor can be expanded as
\begin{align}\label{eq:exchange} 
 {\bf J}_{ij}(Q) \approx {\bf J}_{ij} - \boldsymbol\alpha_{ij} Q,
\end{align}
where the tensor $\boldsymbol\alpha_{ij}$ quantifies the phonon-induced modulation of magnetic interactions.

Within a Holstein--Primakoff expansion, the transverse spin bilinears $S_i^{\alpha}S_j^{\beta}$ with $\alpha,\beta \in \{x,y\}$ contribute terms that are quadratic in the magnon operators (proportional to $a^2$, $a^{\dagger 2}$, or $a^\dagger a$), and the longitudinal term $S_i^z S_j^z$ contributes a term proportional to $a^\dagger a$.
As a consequence, the phonon-modulated exchange generates nonlinear couplings of the form $X_a^2 X_b$, where $X_a = a^\dagger + a$ and $X_b = Q/\ell = b^\dagger + b$, with $\ell = \sqrt{\hbar/(2M\omega)}$ the phonon zero-point length. 
In contrast, a coupling that is \emph{linear} in the magnon operators, and thus capable of opening a direct magnon--phonon hybridization gap, requires a mixed term of the form $S_i^{\alpha} S_j^{z}$ with $\alpha \in \{x,y\}$. 
Such contributions originate from spin--orbit coupling and are therefore suppressed relative to the isotropic exchange $J$ by a factor of order $\alpha_{\mathrm{SOC}} \sim 0.1$.

Restricting the description to the lowest-lying magnon mode and a single phonon mode, and keeping linear and nonlinear interactions up to third order in the boson operators, we obtain the effective Hamiltonian
\begin{align}
 \mathcal{H} &= \omega_{\rm Mag} a^\dagger a + \omega_{\rm Ph_3} b^\dagger b \\
   &+ gX_aX_b+\beta_{aa}^b X_a^2 X_b + \beta_{bb}^a X_b^2 X_a, \nonumber
\end{align}
as presented in the main text.

\noindent\textbf{Simulating the THz 2D spectrum.}
To model the coherent nonlinear dynamics of coupled magnons and phonons in FePS$_3$, we employ a Liouville-space Lindblad framework ~\cite{reitz_nonlinear_2025, Mukamel1986,dorfman_2016}. 
The dynamics of the driven open quantum system 
coupled to the magnetic component $H(t)$ of an external THz field, follows the master equation~\cite{manzano_short_2020,cattaneo_symmtry_2020}
\begin{equation}
    \dot{\rho}(t) = \mathcal{L}\rho(t) + \mathcal{C}_\mu\,H(t)\,\rho(t),
\end{equation}
where the Liouvillian superoperator 
\begin{equation}
    \mathcal{L}\rho = -i[\mathcal{H},\rho] + \sum_j \left(c_j\rho c_j^\dagger - \tfrac{1}{2}\{c_j^\dagger c_j,\rho\}\right),
\end{equation}
combines the coherent evolution of the system under the Hamiltonian $\mathcal{H}$ with dissipation described by the jump operators $\{c_j\}$. 
The light-matter coupling enters through the driving superoperators
\begin{equation}
    \mathcal{C}_\mu\rho = i[\mu,\rho] = i(\mu\rho - \rho\mu),
\end{equation}
where the magnetic field $H(t)$ of the driving laser couples to the system via the magnetic dipole moment operator $\mu = \mu_{\rm drive}$.
In the absence of driving, the formal Liouville-space propagator is $\mathcal{G}(t) = e^{\mathcal{L}t}$.
Vectorizing the density matrix as $|\rho\rrangle = \mathrm{vec}(\rho)$, the equation of motion takes the compact form
\begin{equation}
    \frac{d}{dt}|\rho(t)\rrangle = \mathbf{L}|\rho(t)\rrangle + H(t)\,\mathbf{C}_\mu|\rho(t)\rrangle,
\end{equation}
where $\mathbf{L}$ and $\mathbf{C}_\mu$ denote the sparse matrix representations of the Liouvillian and interaction superoperators, respectively. 
To isolate contributions to different orders in the electric field, we expand the density matrix in powers of the drive as
\begin{equation}
    |\rho(t)\rrangle = \sum_{n=0}^\infty |\rho^{(n)}(t)\rrangle,
\end{equation}
with the initial conditions $\rho^{(0)}(t_0)=\rho_0$ and $\rho^{(n\geq 1)}(t_0)=0$.  
The contribution at the $n$-th order satisfies the recursive equation
\begin{equation}
    \frac{d}{dt}{|\rho^{(n)}(t)\rrangle} = \mathbf{L}|\rho^{(n)}(t)\rrangle + H(t)\,\mathbf{C}_\mu|\rho^{(n-1)}(t)\rrangle.
\end{equation}
The detected signal is given as a linear functional of the density matrix,
\begin{equation}
    S(t) = \mathrm{Tr}[\mu_{\rm det}\rho(t)] = \llangle \mu_{\rm det}^T|\rho(t)\rrangle,
\end{equation}
where $\mu_{\rm det}$ is the detection dipole moment operator of the magnon components, consistent with the experimental $\Delta$LB detection scheme.
The signal at order $n$ in the external field is defined as
\begin{equation}
 S^{(n)}(\tau;t) = \llangle \mu_{\text{det}}^T|\rho^{(n)}(\tau;t)\rrangle.
\end{equation}
Therefore, the second-order signals can be expressed as a double convolution of the driving field with the corresponding response function
\begin{align}
S^{(2)}(t)
= \int_{t_0}^{t}\!dt_1 \int_{t_0}^{t_1}\!dt_2\;
   H(t_1)H(t_2)\,R^{(2)}(t;t_1,t_2),\nonumber \\[3pt]
\end{align}
where the second-order response kernel is defined as
\begin{align}
R^{(2)}(t; t_1, t_2) 
&= \operatorname{Tr}\!\Big[
   \mu_{\rm det}\,\mathcal{G}(t - t_1)\,\mathcal{C}_\mu\, \nonumber\\
&\quad\mathcal{G}(t_1 - t_2)\,\mathcal{C}_\mu\,\rho^{(0)}(t_2)
   \Big].
\end{align}
Similarly, the third-order signal takes the form of a triple field convolution with a three-point response function
\begin{align}
S^{(3)}(t)
&= \int_{t_0}^{t}\! dt_1 
   \int_{t_0}^{t_1}\! dt_2 
   \int_{t_0}^{t_2}\! dt_3\;
   H(t_1) H(t_2) H(t_3)\, \nonumber\\
&\quad R^{(3)}(t; t_1, t_2, t_3)
\end{align}
where the third-order response kernel is given by
\begin{align}
R^{(3)}(t; t_1, t_2, t_3) 
&= \operatorname{Tr}\!\Big[
   \mu_{\rm det}\,\mathcal{G}(t - t_1)\,\mathcal{C}_\mu\, \mathcal{G}(t_1 - t_2)\,\mathcal{C}_\mu\, \nonumber\\
&\hspace{1.0cm}\mathcal{G}(t_2 - t_3)\,\mathcal{C}_\mu\,\rho^{(0)}(t_3)
   \Big].
\end{align}

As illustrated in Fig.~\ref{fig1}(a), we consider two successive laser pulses $H_A(\tau;t)$ and $H_B(t)$, separated by a delay $\tau$, following the standard configuration of 2D THz spectroscopy.~\cite{blank2023two}
The total magnetic field is given by
\begin{equation}
    H_{AB}(\tau;t) = H_A(\tau;t) + H_B(t).
\end{equation}
where $\tau$ denotes the time delay between pulses A and B, and $t$ is the detection time measured from the arrival of pulse B.
In line with the experiments, the simulations employ a broadband driving field spanning the frequency range $0$--$8$~THz.
To disentangle nonlinear responses, we perform three independent simulations: one with both pulses A and B present, and two with either pulse A or pulse B alone.
The nonlinear component is then extracted by subtracting the linear responses,
\begin{equation}
    R^{(n)}_{\text{NL}}(\tau;t) = R^{(n)}_{AB}(\tau;t) - R^{(n)}_{A}(\tau;t) - R^{(n)}_{B}(\tau;t).
    \label{eq:nonlinear_process}
\end{equation}
where $ R^{(n)}_{AB}(\tau;t)$, $R^{(n)}_{A}(\tau;t)$ and $R^{(n)}_{B}(\tau;t)$ denote the $n$-th order responses to two-pulse A and B, single-pulse A and single-pulse B excitation, respectively.
The resulting two-dimensional terahertz spectrum is defined in $(\Omega_\tau,\Omega_t)$, obtained as the Fourier conjugates of the delay $\tau$ and the detection time $t$.
Fourier transformation of Eq.~\ref{eq:nonlinear_process} yields the second- and third-order response amplitudes $S^{(2)}$ and $S^{(3)} $, while contributions beyond the third order are neglected. 
The total spectral amplitude is therefore
\begin{align}  
    S_{\text{total}} &\approx |S^{(2)}+S^{(3)}|.
    \label{eq:spectra_total}
\end{align}
The lineshapes and aspect ratios of the spectral peaks are governed by dissipation and dephasing, as is parameterized in the simulations by the effective decay rates $\kappa$ and $\gamma_\phi$.
All spectra are computed from time-delay scans with strong windowing, yielding a frequency resolution of $\sim 0.033$~THz.
\\

\noindent\textbf{Nonclassical effects.}
In our framework, each bosonic mode (the magnon $a$ and the phonon $b$) is characterized by its canonical quadratures~\cite{serafini_quantum_2023} $X_{a} = a^\dagger + a$ and $P_{a} = i(a^\dagger - a)$, with analogous definitions for mode $ b$. 
The quantum fluctuations of a single mode are encoded in the $2\times2$ covariance matrix 
\begin{equation}
    V = 
\begin{pmatrix}
V_{\rm XX} & C \\
C & V_{\rm PP}
\end{pmatrix},
\label{eq:cov}
\end{equation} 
whose diagonal matrix elements $V_{\rm XX} = \langle X^2\rangle-\langle X\rangle^2$ and $V_{\rm PP} = \langle P^2\rangle-\langle P\rangle^2$ are the quadrature variances, and whose off-diagonal elements $C = \tfrac{1}{2}\mathrm{Tr}[(XP+PX)\rho]-\mathrm{Tr}[X\rho]\mathrm{Tr}[P\rho]$ captures $X$-$P$ correlation.
Diagonalizing the $V$ yields the minimal and maximal variances of the mode. 
For dimensionless quadratures satisfying the canonical commutation relation $[X,P] = 2i$, the single-mode squeezing is identified by $V_{\min} < 1$.
Relative to the vacuum noise level $V_0 = 1$, the squeezing level in decibel (dB) units is given by $S_{\mathrm{dB}} = 10\log_{10}V_{\min}$, with negative values indicating the noise suppression below the vacuum limit.~\cite{wolfgang_quantum_2001}

The single-mode analysis extends naturally to the two-mode system.
Collecting the canonical variables in the quadrature vector $\mathbf{R} = (X_a, P_a, X_b, P_b)^T$, the covariance matrix elements are defined as
$V_{ij} = \frac{1}{2}\langle R_i R_j + R_j R_i \rangle-\langle R_i \rangle \langle R_j \rangle$, with the fluctuation operators $\Delta X_a = X_a - \langle X_a\rangle$, $\Delta P_a = P_a - \langle P_a\rangle$ (and analogously for mode $b$).
The resulting two-mode covariance matrix for the coupled magnon-phonon system reads
\begin{equation}
\resizebox{0.9\hsize}{!}{$
V =
\begin{pmatrix}
\langle \Delta X_a^{2} \rangle
& \tfrac12\langle \Delta X_a\Delta P_a+\Delta P_a\Delta X_a\rangle
& \tfrac12\langle \Delta X_a\Delta X_b+\Delta X_b\Delta X_a\rangle
& \tfrac12\langle \Delta X_a\Delta P_b+\Delta P_b\Delta X_a\rangle
\\
\tfrac12\langle \Delta P_a\Delta X_a+\Delta X_a\Delta P_a\rangle
& \langle \Delta P_a^{2}\rangle
& \tfrac12\langle \Delta P_a\Delta X_b+\Delta X_b\Delta P_a\rangle
& \tfrac12\langle \Delta P_a\Delta P_b+\Delta P_b\Delta P_a\rangle
\\
\tfrac12\langle \Delta X_b\Delta X_a+\Delta X_a\Delta X_b\rangle
& \tfrac12\langle \Delta X_b\Delta P_a+\Delta P_a\Delta X_b\rangle
& \langle \Delta X_b^{2}\rangle
& \tfrac12\langle \Delta X_b\Delta P_b+\Delta P_b\Delta X_b\rangle
\\
\tfrac12\langle \Delta P_b\Delta X_a+\Delta X_a\Delta P_b\rangle
& \tfrac12\langle \Delta P_b\Delta P_a+\Delta P_a\Delta P_b\rangle
& \tfrac12\langle \Delta P_b\Delta X_b+\Delta X_b\Delta P_b\rangle
& \langle \Delta P_b^{2}\rangle
\end{pmatrix}.
\label{eq:V_matrix}
$}
\end{equation}

To quantify the magnon-phonon entanglement, we apply a partial transpose $V^{\rm PT} = \Lambda V \Lambda$ to the covariance matrix with $\Lambda = \mathrm{diag}(1,1,1,-1)$, and compute its symplectic eigenvalues $\tilde{\nu}_{\pm}$.
To quantify the magnon-phonon entanglement, we perform a partial transpose $V^{\rm PT} = \Lambda V \Lambda$ on the covariance matrix, with $\Lambda = \mathrm{diag}(1,1,1,-1)$, and compute its symplectic eigenvalues $\tilde{\nu}_{\pm}$.
According to the Peres--Horodecki criterion, also known as the Positive Partial Transpose (PPT) criterion,~\cite{peres_separability_1996,horodecki_inseparable_1997,simon_peres-horodecki_2000} the two modes are entangled whenever the smallest symplectic eigenvalue of the partially transposed covariance matrix satisfies $\tilde{\nu}_{\min} < 1$.
The entanglement strength is then quantified by the logarithmic negativity,~\cite{Plenio_logarithmic_2005}
$ E_{\mathcal N}=\max\!\left[0, -\ln \tilde{\nu}_{\min}\right]$.
To further quantify the two-mode squeezing in this asymmetric system, we evaluate the Duan inseparability criterion through the generalized EPR operators~\cite{duan_inseparability_2000} 
\begin{equation}
    u=|g|X_{a}-X_{b}/g, \quad
    {v}=|g|{P}_{a}+{P}_{b}/g.
\end{equation}
Because the magnon-phonon system is asymmetric, the optimal scaling parameter g$_{\mathrm{opt}}$ is determined from the covariance matrix in Eq.~\ref{eq:V_matrix}.
Inserting $g_{\mathrm{opt}}$ into the Duan criterion, two-mode EPR squeezing occurs when the combined variance satisfies $\mathrm{Var}(u)+\mathrm{Var}(v) < g^{2}+1/g^{2}$. 
In addition, the maximally squeezed collective quadrature is identified by performing a rotation in the two-mode quadrature space,
\begin{equation}
\begin{pmatrix}
\tilde X_1 \\[4pt]
\tilde X_2
\end{pmatrix}
=
R_X
\begin{pmatrix}
X_a \\[4pt]
X_b
\end{pmatrix},
\qquad
R_X =
\begin{pmatrix}
C_a & C_b \\
-\,C_b & C_a
\end{pmatrix},
\label{eq:RX}
\end{equation}
with $C_a^2 + C_b^2 = 1$. 
This defines the rotated collective quadratures $\tilde X_1$ and $\tilde X_2$.
Diagonalizing the $X$-quadrature matrix $V_X$, the eigenvalues $V_{X,\min}$ and $V_{X,\max}$ are obtained, corresponding to the most squeezed and most anti-squeezed collective modes. 
Optimal two-mode squeezing is present when
\begin{equation}
V_{X,\min} < 1,
\end{equation}
where the vacuum noise level is $1$.



\section{Acknowledgements}
The authors thank Qian Cheng and Zizhan Gao for their assistance in optical measurements. The authors also thank Zihang Song, Dr. Song Bao and Dr. Jinsheng Wen for their assistance in Laue measurements. This work was supported by the National Key Research and Development Program of China (Grant Nos.~2020YFA0309200, 2024YFA1408700, and 2022YFA1403800), the National Natural Science Foundation of China (Nos.~12474475, 12574349, 92477115, 12250008, and 12188101), the Natural Science Foundation of Jiangsu Province (Nos. BK20240057, BK20243011, BK20233001), and the Fundamental Research Funds for the Central Universities (Nos.~14380184). N.W., E.V.B., and A.R. acknowledge support from the Cluster of Excellence “CUI: Advanced Imaging of Matter”-EXC 2056–project ID 390715994, SFB-925 “Light induced dynamics and control of correlated quantum systems”–project ID 170620586 of the Deutsche Forschungsgemeinschaft (DFG), the European Research Council (ERC-2024-SyG-UnMySt–101167294), and the Max Planck-New York City Center for Non-Equilibrium Quantum Phenomena. The Flatiron Institute is a division of the Simons Foundation. F. Liu thanks the support from Sichuan Province Key Laboratory of Display Science and Technology. A portion of this work was supported by the Synergetic Extreme Condition User Facility (SECUF, https://cstr.cn/31123.02.SECUF).

\section{Author Contributions}
Q.Z. conceived the project.
Z.Li grew the single crystals under the supervision of F.L.;
L.Li performed 1D and 2D THz-pump optical-probe measurements under zero field and analyzed the data under the supervision of Q.Z.;
L.Li and Z.Lin performed measurements under magnetic fields in SECUF, supported by X.W. and J.L.;
N.W., E.V.B., and A.R. provided the theoretical modeling and simulations.
L.Liu and Y.W. contributed to data analysis and interpretation.
L.Li, Q.Z., N.W., and E.V.B. wrote the manuscript with critical input from all other authors.

\section{Data availability\label{sec:data availability}}
All data that support the plots within this paper and other findings of this study are available upon request.

\section{Additional Information}
The authors declare no competing interests.

\clearpage
\setcounter{figure}{0} 
\renewcommand{\figurename}{Extended Data Figure}
\section{Extended Data Figures}

\begin{figure*}[h]
    \centering
	\includegraphics[width=1\textwidth]{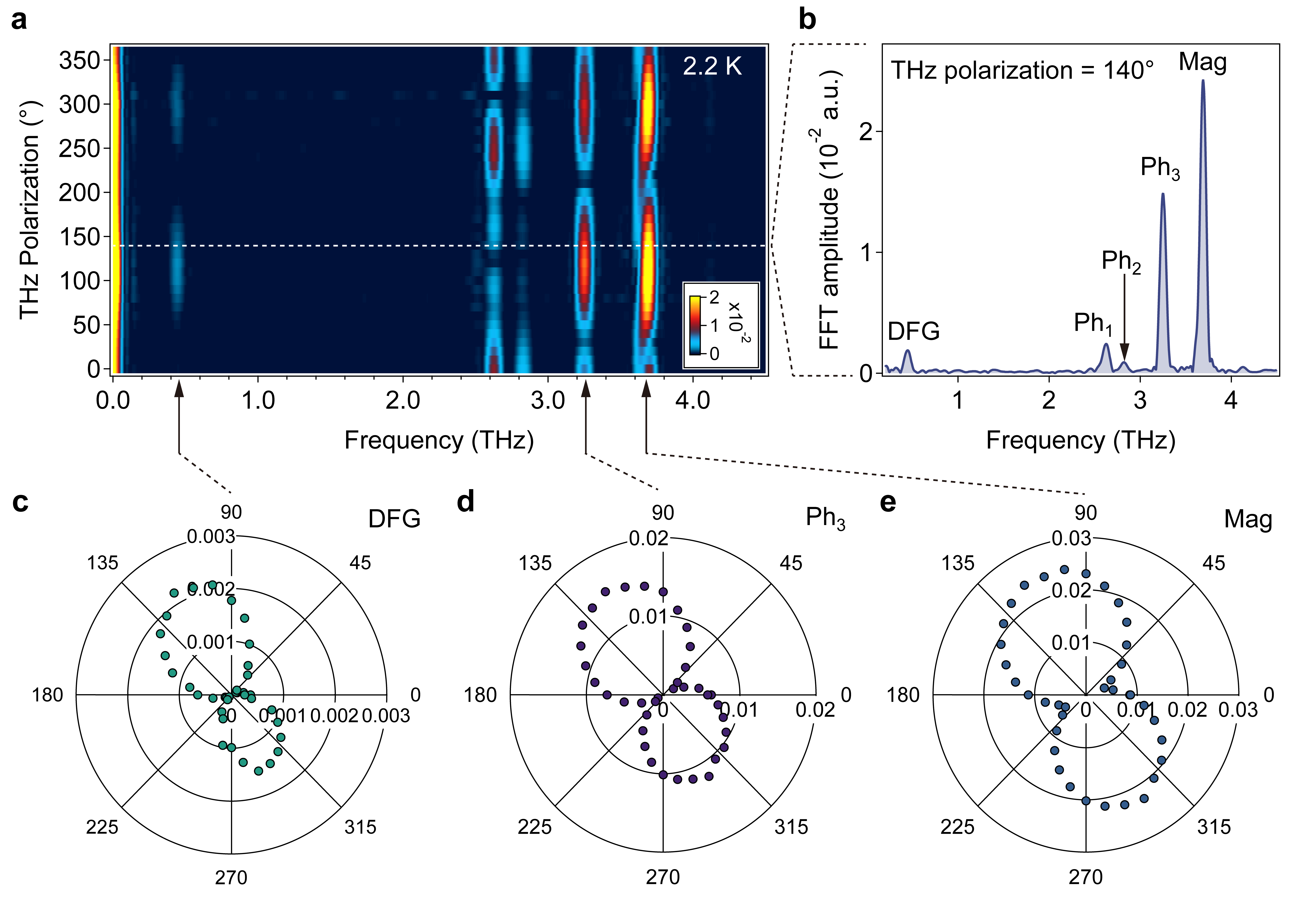}
	\caption{\textbf{Polarization dependence of THz-driven coherent AFM magnon and phonons in FePS$_{3}$.} \textbf{a},~Polarization dependence of Mag-Ph$_{3}$ DFG, Ph$_{1}$, Ph$_{2}$, Ph$_{3}$ and Mag signals. \textbf{b}, The frequency-domain spectrum with 140° THz electric field polarization, corresponding to the polarization setting described in the main text. \textbf{c–e}, THz polarization dependence of Mag-Ph$_{3}$ DFG and Ph$_{3}$ and Mag modes. Intense THz pulses with 2.5 MV/cm electric field strength were used in the measurements.}
\label{S:1}
\end{figure*}
\clearpage

\begin{figure*}[h]
    \centering
	\includegraphics[width=1\textwidth]{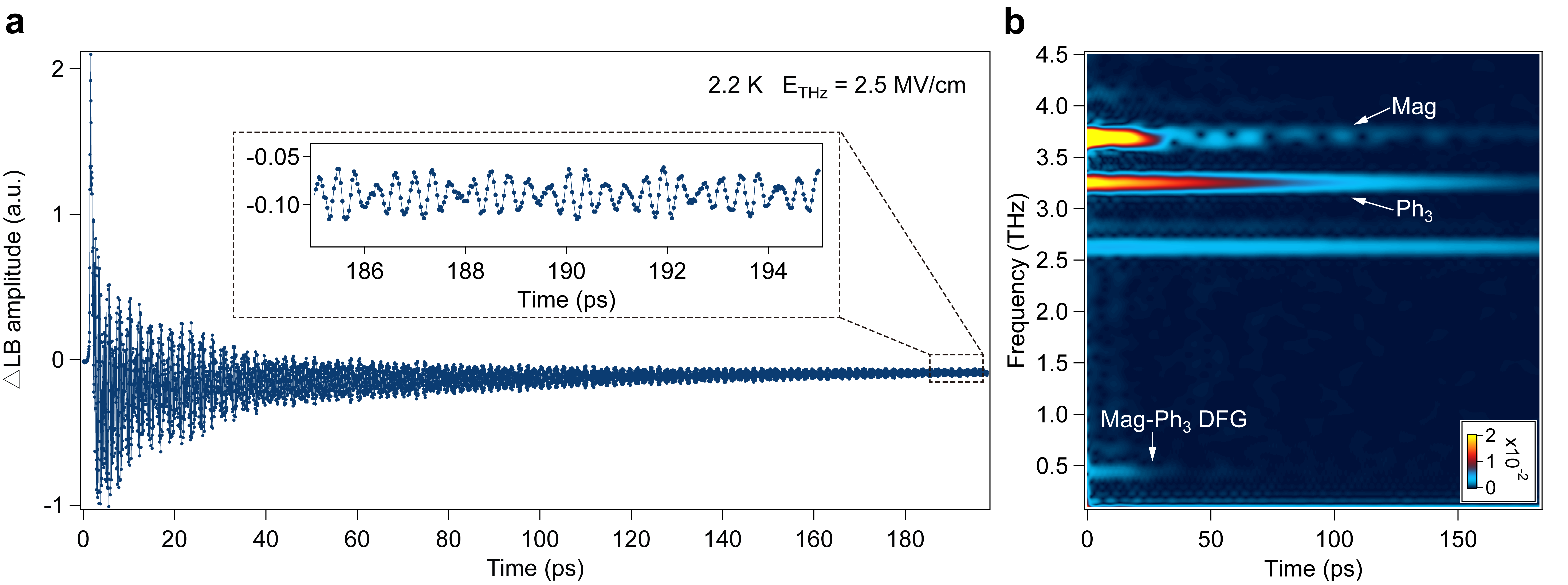}
	\caption{\textbf{Long-lived coherent magnons and phonons induced by intense THz pulses with 2.5 MV/cm electric field.} \textbf{a}, Time-domain coherent oscillations observed up to 200~ps. \textbf{b}, Spectrogram of the time-domain oscillation in \textbf{a}. The time window for the wavelet transform is 15~ps.}
\label{S:2}
\end{figure*}

\newpage
\begin{figure*}[h]
    \centering
	\includegraphics[width=1\textwidth]{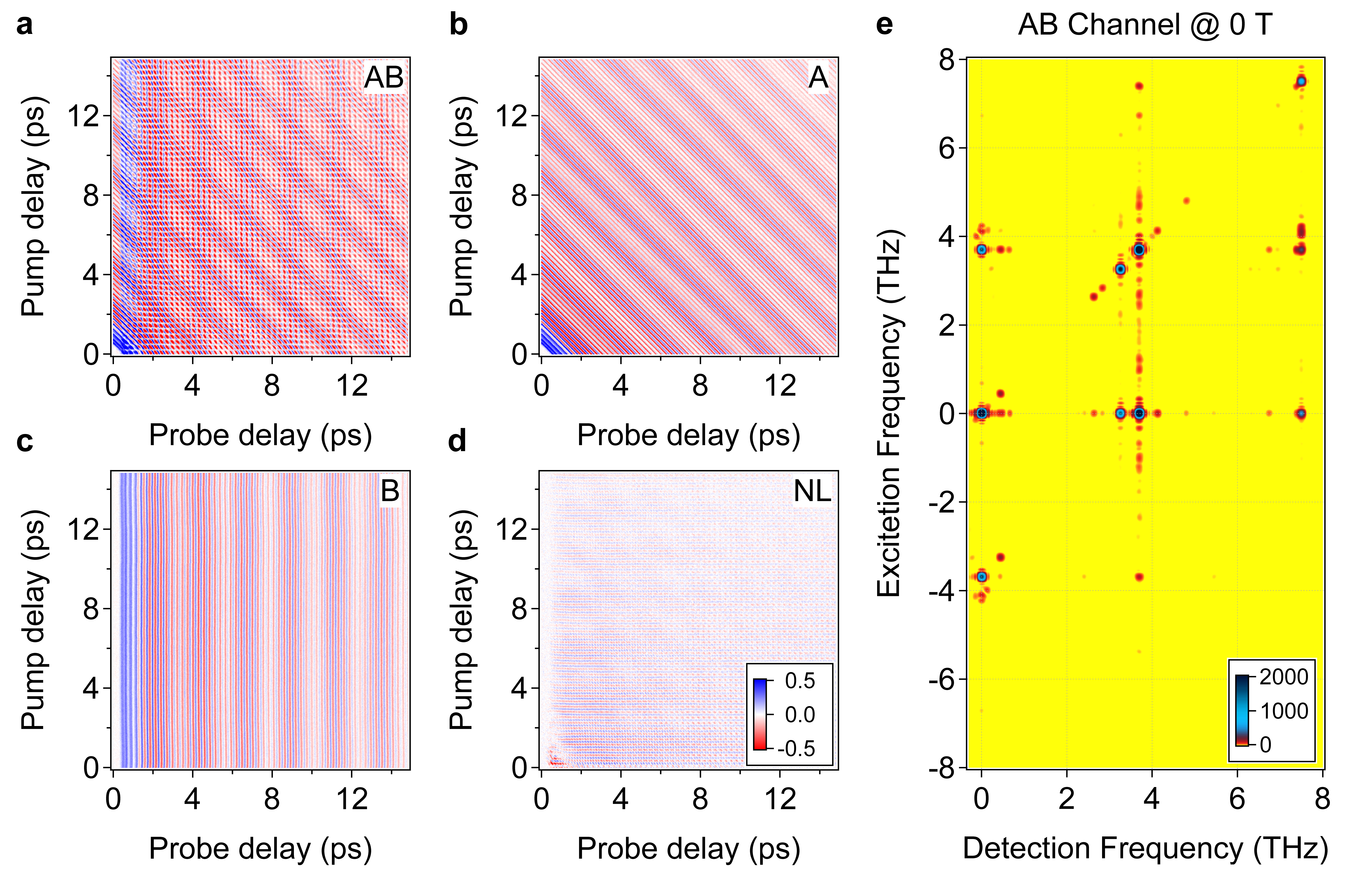}
	\caption{\textbf{Time-domain 2D spectra of FePS$_{3}$ at 0~T.} \textbf{a–d},~The 2D time domain signal with 0.033 ps scan step for both pump delay and probe delay in AB (\textbf{a}), A (\textbf{b}), B (\textbf{c}), NL (\textbf{d}) channels. The signals in the NL channel are induced by both A and B THz pulses, which correspond to the frequency-domain spectrum shown in Fig.~\ref{fig2}a. \textbf{e}, The 2D frequency-domain spectrum of the AB channel, which contains both linear and nonlinear signals.}
\label{S:3}
\end{figure*}

\newpage
\begin{figure*}[h]
    \centering
	\includegraphics[width=1\textwidth]{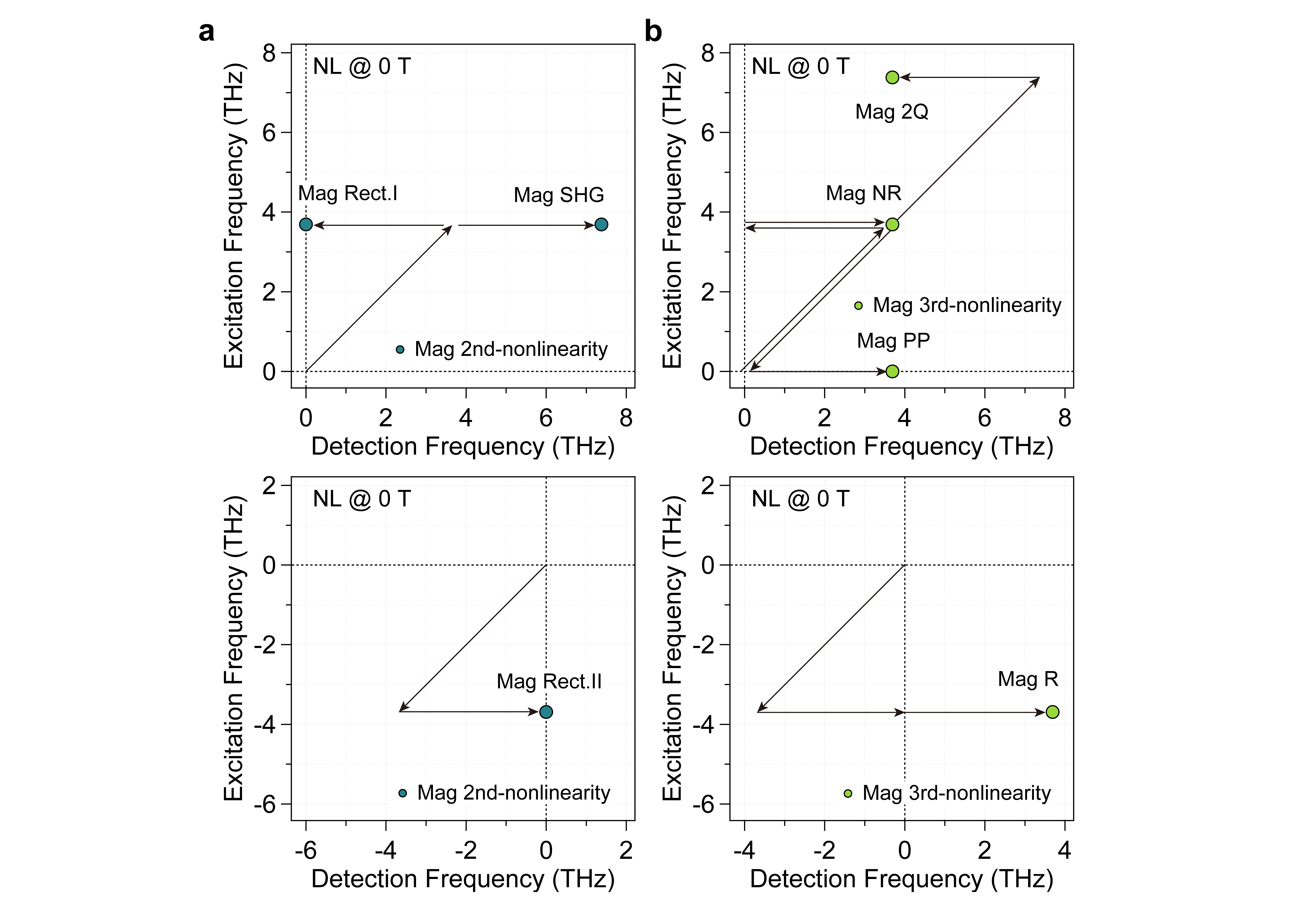}
	\caption{\textbf{Frequency-vector diagram of nonlinear magnonic processes at 0 T.} \textbf{a}, The second-order magnonic processes, Magnon Rectification and SHG. \textbf{b}, The third-order nonlinear processes, magnon 2Q, NR, PP and R. The arrows in the diagonal direction correspond to the excitation from pump pulse A, while the horizontal arrows are from pump pulse B.}
\label{S:4}
\end{figure*}

\newpage
\begin{figure*}[h]
    \centering
	\includegraphics[width=1\textwidth]{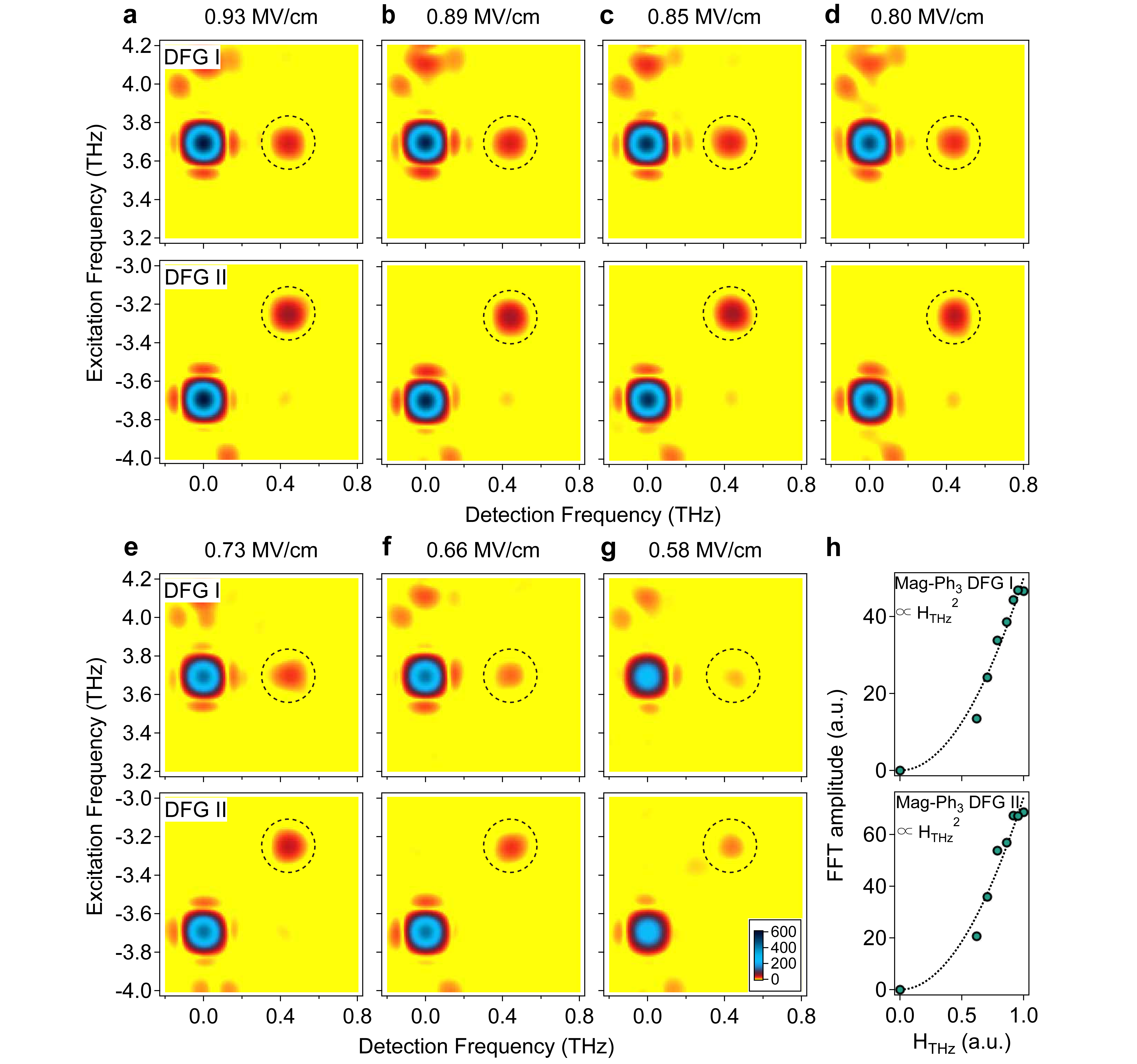}
	\caption{\textbf{The THz-field dependence of Mag-Ph$_{3}$ DFG signal in 2D spectrum at 0~T.} \textbf{a–g}, Mag-Ph$_{3}$ DFG raw spectra in DFG I/II region pumped by different $E_\text{THz}$. \textbf{h}, Field dependence of Mag-Ph$_{3}$ DFG signal amplitude shows the quadratic scaling in a limited amount of data. }
\label{S:5}
\end{figure*}

\newpage
\begin{figure*}[h]
    \centering
	\includegraphics[width=1\textwidth]{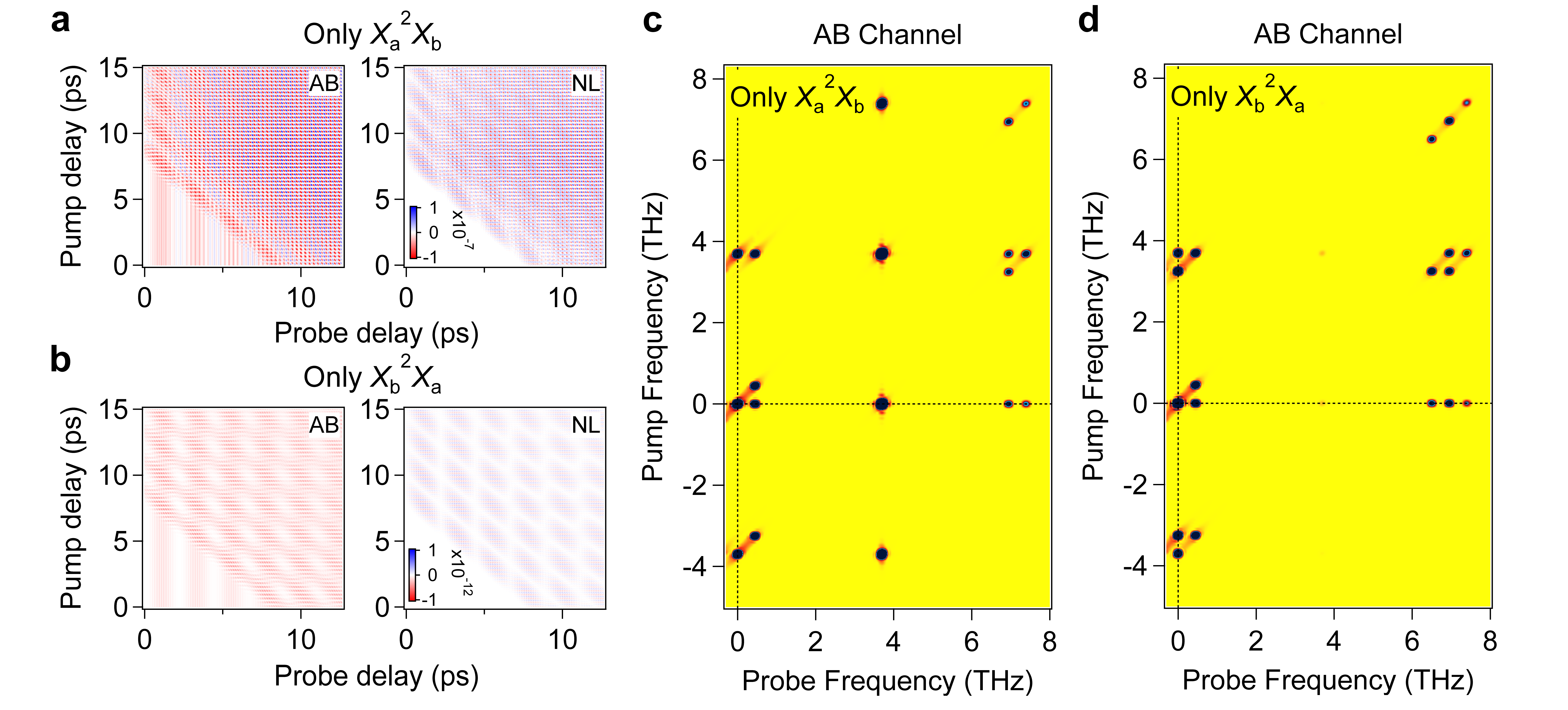}
	\caption{\textbf{Calculated time-domain 2D spectra and corresponding FFT results.} \textbf{a, b}, Two-dimensional time-domain signal of the AB and NL channel calculated with three-wave mixing term $X_a^2X_b$ or $X_b^2X_a$. \textbf{c, d}, The corresponding frequency-domain spectra of the AB channel. The associated NL channel signals are shown in Fig.~\ref{fig4}.}
\label{S:6}
\end{figure*}

\newpage
\begin{figure*}[h]
    \centering
	\includegraphics[width=0.95\textwidth]{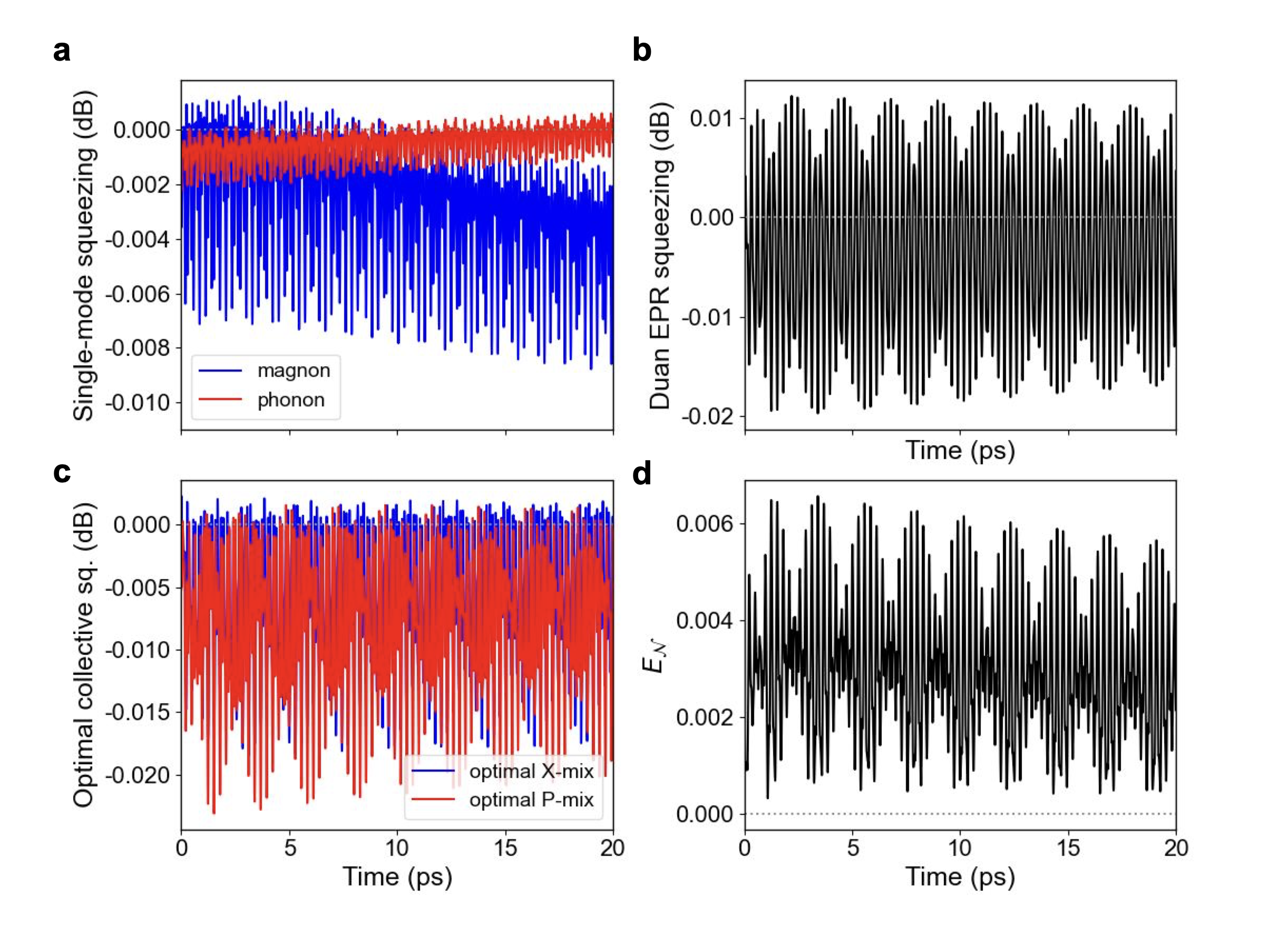}
	\caption{\textbf{Calculated squeezing and entanglement of magnon and Ph$_{3}$ phonon induced by an intense THz pulse.} \textbf{a}, Single-mode squeezing of magnon and Ph$_{3}$ phonons as a function of time. \textbf{b} Time-dependent two-mode EPR squeezing and \textbf{c} two-mode optimal squeezing. \textbf{d}, Degree of entanglement between magnon and Ph$_{3}$ phonons. In the calculation, the nonlinear coupling term is $\beta_{aa}^bX_a^2X_b$, with $\beta_{aa}^b$ = 17~meV/\AA. The excitation THz pulse has a double-cycled Gaussian profile with a field strength of 2~MV/cm. See Methods for calculation details.}
    
\label{S:7}
\end{figure*}

\end{document}